\documentclass[twocolumn,showpacs,preprintnumbers,amsmath,amssymb]{revtex4}

\usepackage{graphicx}
\usepackage{dcolumn}
\usepackage{bm}

\begin{document}


\title{Projection Measurement of the Maximally Entangled N-Photon State
\\for a Demonstration of N-Photon de Broglie Wavelength}

\author{F. W. Sun$^1$, Z. Y. Ou$^{1,2}\footnote{E-mail: zou@iupui.edu}$, and G. C. Guo$^1$}
 \affiliation{$^1$Key Laboratory of Quantum Information,
 University of Science and Technology of China, CAS, Hefei, 230026, the People's Republic of China
 \\$^2$Department of Physics, Indiana
University-Purdue University Indianapolis \\ 402 N. Blackford
Street, Indianapolis, IN 46202}

\date{\today}

\begin{abstract}
We construct a projection measurement process for the maximally
entangled N-photon state (the NOON-state) with only linear optical
elements and photodetectors. This measurement process will give
null result for any N-photon state that is orthogonal to the NOON
state. We examine the projection process in more detail for $N=4$
by applying it to a four-photon state from type-II parametric
down-conversion. This demonstrates an orthogonal projection
measurement with a null result. This null result corresponds to a
dip in a generalized Hong-Ou-Mandel interferometer for four
photons. We find that the depth of the dip in this arrangement can
be used to distinguish a genuine entangled four-photon state from
two separate pairs of photons. We next apply the NOON state
projection measurement to a four-photon superposition state from
two perpendicularly oriented type-I parametric down-conversion
processes. A successful NOON state projection is demonstrated with
the appearance of the four-photon de Broglie wavelength in the
interference fringe pattern.
\end{abstract}

\maketitle

\section{\label{sec:level1}Introduction}

Recently, attentions
\cite{bol,ou,lee,kok,fiu,hof,wal,mit,sso,wang} have been focussed
on the generation of the so-called NOON state in the form of
\begin{eqnarray}
|NOON\rangle = {1\over \sqrt{2}}(|N,0\rangle + |0,
N\rangle).\label{NOON}
\end{eqnarray}
It has been shown \cite{bol,ou} that the NOON state has the
advantage in the sensitivity of optical interferometry over a
coherent state and can achieve the Heisenberg limit \cite{hei} of
$1/N$ in the accuracy of phase measurement. This is a factor of
$\sqrt{N}$ improvement over the traditional standard quantum limit
of $1/\sqrt{N}$ with a coherent state. Experimental demonstration
for $N=2$ was first performed by Ou et al \cite{zou} and by Rarity
et al \cite{rar}, and more recently by Fonseca et al \cite{fon}
and by Edamatsu et al \cite{eda}. The extension to $N=3$ and 4 was
reported by Mitchell et al \cite{mit} and Walther et al
\cite{wal}. Boto et al \cite{bot} have shown that quantum
lithography with the NOON state can increase the resolution by a
factor of $N$ compared to a coherent state. This is so because the
NOON state shows a de Broglie wavelength of $\lambda/N$ for the
$N$-photon interference. D'Angelo et al \cite{shi} demonstrated
the feasibility of the scheme for $N=2$.

The general trend in preparing the NOON state is by the method of
quantum interference \cite{hom,hof,sso,wang} for the cancellation
of the unwanted states of $|N-1,1\rangle, |N-2,2\rangle, ... $ ,
etc. A simplest example is the Hong-Ou-Mandel interferometer
\cite{hom} where two photons enter a 50:50 lossless beam splitter
from two input sides separately ($|\Phi_2\rangle_{in} =
|1,1\rangle$). The output state is a NOON state of $N=2$:
\begin{eqnarray}
|\Phi_2\rangle_{out} = {1\over \sqrt{2}}\Big(|2,0\rangle + |0,
2\rangle\Big).\label{HOM}
\end{eqnarray}
The disappearance of the $|1,1\rangle$ in the output state in
Eq.(\ref{HOM}) is a result of quantum two-photon interference.
Unfortunately, extension to four-photon state of
$|\Phi_4\rangle_{in} = |2,2\rangle$ will not produce a four-photon
NOON state but a state of \cite{rhe}
\begin{eqnarray}
|\Phi_4\rangle_{out} = \sqrt{{3\over 8}}\Big(|4,0\rangle + |0,
4\rangle\Big)+{1\over 2}|2,2\rangle.\label{HOM2}
\end{eqnarray}
It is impossible to take out the unwanted $|2,2\rangle$ state with
just linear optical elements by quantum interference. It is
important that our starting state is in the form of $|2,2\rangle$
because it is what we can have from a parametric down-conversion
process.

In this paper, we approach this problem from another aspect, that
is, the measurement process. If our measurement only responds to
the first term, i.e., the NOON state, but not to the second term,
i.e., $|2,2\rangle$ in Eq.(\ref{HOM2}), we will not have the
contribution from the $|2,2\rangle$ term and we effectively obtain
the four-photon NOON state.

Recently, an issue has been raised about how to distinguish an
entangled multi-photon state from a quantum state with photons
well separated and distinguishable \cite{tsu,ou2}. The difference
lies in the multi-photon interference: an entangled multi-photon
state will give rise to the strongest multi-photon interference
effect whereas a distinguishable multi-photon state produce less
or sometimes zero interference effect. It turns out that the new
projection measurement scheme that we are going to introduce is
based on a multi-photon interference effect to cancel the
contributions from the orthogonal states. Thus this scheme will be
able to quantitatively characterize the degree of entanglement for
a multi-photon state.

In the following, we will first introduce the general NOON state
projection measurement. We then apply it to the $|2,2\rangle$
state which is orthogonal to the four-photon NOON state. We will
show that we obtain a null result confirming the orthogonality.
Next we apply the projection measurement to the state in
Eq.(\ref{HOM2}) and project out the NOON state for a demonstration
of four-photon de Broglie wavelength. We conclude with a
discussion.

\section{NOON State Projection Measurement}

The general idea in constructing a NOON state projection
measurement is very similar to Hofmann's method \cite{hof} of
creating the NOON state by super-bunching of N independent
photons. It follows from the algebraic identity:
\begin{eqnarray}
x^N-y^N = \prod_{n=0}^{N-1}(x-ye^{i2n\pi/N}).\label{factor}
\end{eqnarray}
If we substitute $x, y$ by $\hat a_H, \hat a_V$ with $H,V$
representing the horizontal and vertical polarizations, we have
the same equation but for the operators
\begin{eqnarray}
\hat a_H^N-\hat a_V^N = \prod_{n=0}^{N-1}(\hat a_H-\hat
a_Ve^{i2n\pi/N}).\label{op-factor}
\end{eqnarray}
This is possible because $\hat a_H$ and $\hat a_V$ commute. Next,
we make a joint measurement of the operators $\hat
b_n^{\dagger}\hat b_n$ with
\begin{eqnarray}
\hat b_n\equiv (\hat a_H-\hat a_Ve^{i2n\pi/N})/\sqrt{2}~~~~~(n=0,
...,N-1),\label{bn}
\end{eqnarray}
the joint probability $P_N$ is proportional to
\begin{eqnarray}
&&P_N\propto \langle\Phi_N|\hat b_{N-1}^{\dagger}...\hat
b_0^{\dagger}\hat b_0 ... \hat b_{N-1}|\Phi_N\rangle\cr &&\hskip
0.24in \propto \langle\Phi_N|\hat a_H^{\dagger N}- \hat
a_V^{\dagger N})(\hat a_H^{N}-\hat
a_V^{N})|\Phi_N\rangle,\label{P}
\end{eqnarray}
where
\begin{eqnarray}
|\Phi_N\rangle = \sum_{n=0}^{N}c_n|N-n, n\rangle \label{Ph}
\end{eqnarray}
is an arbitrary N-photon state for two modes of $\hat a_H$ and
$\hat a_V$. It is easy to see that only terms with $n=0,N$, i.e.,
only the NOON state part in the general state in Eq.(\ref{Ph})
contribute to $P_N$ in Eq.(\ref{P}). Thus we achieve a NOON state
projection measurement and
\begin{eqnarray}
P_N\propto |c_0-c_N|^2.\label{PN}
\end{eqnarray}
If we introduce a phase shift between $\hat a_H$ and $\hat a_V$,
i.e., $\varphi = \varphi_H-\varphi_V$, $c_0^*c_N = |c_0
c_N|e^{iN\varphi}$ and Eq.(\ref{PN}) becomes
\begin{eqnarray}
P_N\propto |c_0|^2 + |c_N|^2 -2|c_0 c_N|\cos N\varphi.\label{PN2}
\end{eqnarray}
Here $\varphi$ is single photon phase difference. For the special
case of $|c_0|=|c_N|$, we have
\begin{eqnarray}
P_N\propto 1-\cos N\varphi.\label{PN3}
\end{eqnarray}
The dependence on $N\varphi$ is a signature of N-photon de Broglie
wave.

\begin{figure}[htb]
\begin{center}
\includegraphics[width= 3.3in]{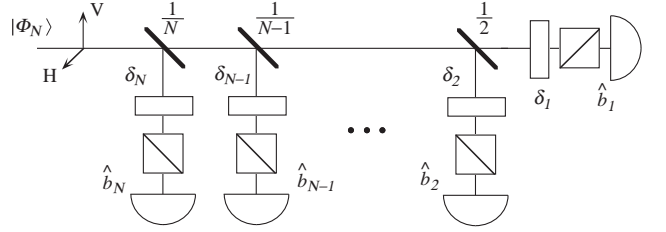}
\end{center}
\caption{\em Layout for the NOON state projection measurement. The
number above each beam splitter is the reflectivity. $\delta_k =
2(k-1)\pi/N$ is the phase delay between H and V polarizations. The
polarizers are 45 degree oriented.} \label{fig1}
\end{figure}

It is straightforward to achieve the projection measurement
discussed above. Consider the scheme depicted in Fig.1. It is easy
to show that the operator in front of each detector has the form
given in Eq.(\ref{bn}). Thus the n-th detector measures $\hat
b_n^{\dagger}\hat b_n$. The N-fold coincidence measurement of all
N detectors will give $P_N$, the outcome of a NOON state
projection measurement. Notice that state projection methods have
been proposed before \cite{lee,kok,fiu}. But those methods require
$2N$ photons for the generation of $N$-photon NOON state.

For a special case of N=4, the actual implementation of the NOON
state measurement is shown in Fig.2. The operators of the four
detectors have the following form
\begin{eqnarray}
\begin{cases}
\hat b_1= (\hat a_H-\hat a_V)/2 + \hat b_{01}\cr \hat b_2= (\hat
a_H+\hat a_V)/2+ \hat b_{02}\cr\hat b_3= (\hat a_H-i\hat a_V)/2+
\hat b_{03}\cr\hat b_4= (\hat a_H+i\hat a_V)/2+ \hat b_{04}
\end{cases} \label{b4}
\end{eqnarray}
Here $\hat b_{0n}~(n=1-4)$ are some operators related to the
vacuum modes $\hat a_{0H,V}$ in the unused beam splitter input
port and make no contribution to photon detection.

\begin{figure}[htb]
\begin{center}
\includegraphics[width= 2.5in]{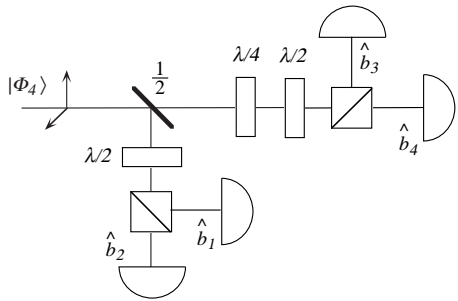}
\end{center}
\caption{\em NOON state projection measurement for N=4.
$\lambda/2$ is a half wave plate for 45 degree polarization
rotation whereas $\lambda/4$ is a quarter wave plate.}
\label{fig2}
\end{figure}

\section{Orthogonal Projection}

Let us now consider an input state of $|2,2\rangle$ for the NOON
state projection measurement. This state can be produced in
nondegenerate parametric down-conversion (NPDC). According to the
previous discussion, we should have $P_4(|2,2\rangle)=0$ because
$|2,2\rangle$ is orthogonal to the NOON state. In practice,
however, we do not exactly have the state $|2,2\rangle$ in NPDC.

\subsection{Simple Pictures for Two Independent Pairs and Four
Entangled Photons}

Usually in parametric down-conversion, a pair of photons is
generated with extremely short correlation time ($T_c \sim 100$
fs) between the two photons but different pairs are produced
completely in random as shown in Fig.3a. This case of well
separated distinguishable two pairs of photons is called a
$2\times 2$ case. On the other hand, when ultra short pump pulses
are applied to ensure the two pairs be produced within the short
pump pulse duration, the four photons form an indistinguishable
four-photon entangled state, as in Fig.3b. This case is called a
$4\times 1$ case. Sometimes, the paths between the two correlated
photons (within one pair of photons) may not be balanced. Then all
four photons are well separated and we call this situation a
$1\times 4$ case, as shown in Fig.3c.

\begin{figure}[htb]
\begin{center}
\includegraphics[width= 3.0in]{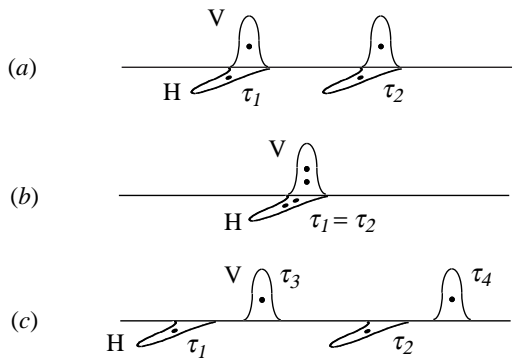}
\end{center}
\caption{\em Two pairs of photons from a Type-II parametric
down-conversion: (a) two pairs are not generated in the same time
and are distinguishable in time (the $2\times 2$ case); (b) two
pairs are indistinguishable (the $4\times 1$ case); (c) all four
photons are distinguishable (the $1\times 4$ case).} \label{fig3}
\end{figure}

Let's now label the times at which the two pairs are generated as
$\tau_1, \tau_2$, respectively. For the case in Fig.3a ($2\times
2$ case), we have $|\tau_1-\tau_2|>> T_c$ but for Fig.3b ($4\times
1$ case), $|\tau_1-\tau_2|<< T_c$. We can then write the quantum
state of the two pairs as
\begin{eqnarray}
|\Phi\rangle = |\phi(\tau_1)\rangle \otimes
|\phi(\tau_2)\rangle,\label{phi-tau}
\end{eqnarray}
with
\begin{eqnarray}
|\phi\rangle = |H V\rangle .
\end{eqnarray}

Consider the detection scheme in Fig.2 for $N=4$ case. The
four-photon coincidence is proportional to the four-photon
correlation function:
\begin{eqnarray}
&&G^{(4)}(t_1,t_2,t_3,t_4) \cr&&\hskip 0.3 in = ||\hat
E_1(t_1)\hat E_2(t_2)\hat E_3(t_3)\hat
E_4(t_4)|\Phi\rangle||^2,\label{G4hv}
\end{eqnarray}
where
\begin{eqnarray}
\begin{cases}\hat E_1(t) = \big[\hat E_H(t)+\hat
E_V(t)\big]/\sqrt{2}, \cr\hat E_2(t) = \big[\hat E_H(t)-\hat
E_V(t)\big]/\sqrt{2},\cr \hat E_3(t) = \big[\hat E_H(t)+i\hat
E_V(t)\big]/\sqrt{2},\cr \hat E_4(t) = \big[\hat E_H(t)-i\hat
E_V(t)\big]/\sqrt{2}, \end{cases}\label{Ehv}
\end{eqnarray}
where $\hat E_{H,V}(t) = \hat a_{H,V}(t)$ for the single mode
treatment. When applying the above to the state in
Eq.(\ref{phi-tau}), we find six nonzero terms in Eq.(\ref{G4hv}),
that is, when (i) $t_1 =t_2=\tau_1, t_3=t_4=\tau_2$ or $t_1
=t_2=\tau_2, t_3=t_4=\tau_1$; (ii) $t_1 =t_3=\tau_1,
t_2=t_4=\tau_2$ or $t_1 =t_3=\tau_2, t_2=t_4=\tau_1$; (iii) $t_1
=t_4=\tau_1, t_2=t_3=\tau_2$ or $t_1 =t_4=\tau_2, t_2=t_3=\tau_1$.
The case (i) can be calculated as
\begin{eqnarray}
&&\hat E_1(\tau_1)\hat E_2(\tau_1)\hat E_3(\tau_2)\hat
E_4(\tau_2)|\Phi\rangle  \cr &&\hskip 0.2 in = \hat
E_1(\tau_1)\hat E_2(\tau_1)|\phi(\tau_1)\rangle\otimes \hat
E_3(\tau_2)\hat E_4(\tau_2)|\phi(\tau_2)\rangle\cr &&\hskip 0.2 in
= (1/2)^2(1-1)(i-i)|0\rangle = 0. \label{E1}
\end{eqnarray}
It is the same for $\hat E_1(\tau_2)\hat E_2(\tau_2)\hat
E_3(\tau_1)\hat E_4(\tau_1)|\Phi\rangle$. The zero result in this
case stems from the two-photon Hong-Ou-Mandel effect between $E_1$
and $E_2$ and between $E_3$ and $E_4$. Similarly, the case (ii)
gives
\begin{eqnarray}
&&\hat E_1(\tau_1)\hat E_2(\tau_2)\hat E_3(\tau_1)\hat
E_4(\tau_2)|\Phi\rangle \cr &&\hskip 0.3 in = \hat E_1(\tau_2)\hat
E_2(\tau_1)\hat E_3(\tau_2)\hat E_4(\tau_1)|\Phi\rangle\cr
&&\hskip 0.3 in = (1/2)^2 (1+i)(1+i)|0\rangle;\label{E2}
\end{eqnarray}
and the case (iii) gives
\begin{eqnarray}
&&\hat E_1(\tau_1)\hat E_2(\tau_2)\hat E_3(\tau_2)\hat
E_4(\tau_1)|\Phi\rangle \cr &&\hskip 0.3 in =\hat E_1(\tau_2)\hat
E_2(\tau_1)\hat E_3(\tau_1)\hat E_4(\tau_2)|\Phi\rangle \cr
&&\hskip 0.3 in = (1/2)^2(1-i)(1-i)|0\rangle.\label{E3}
\end{eqnarray}

When the two pairs are separated, i.e., $|\tau_1-\tau_2|>> T_c$,
all six contributions are distinguishable ($2\times2$ case) and we
add their absolute values to give $G^{(4)}$:
\begin{eqnarray}
G^{(4)}(2\times2) =  2\times 0 + 2\bigg|{1+i\over 2}\bigg|^4 +
2\bigg|{1-i\over 2}\bigg|^4 = 1 .\label{GG4}
\end{eqnarray}
On the other hand, when the two pairs overlap and become
indistinguishable, i.e., $|\tau_1-\tau_2|<< T_c$ ($4\times 1$
case), we add the six amplitudes:
\begin{eqnarray}
G^{(4)}(4\times 1)\propto \bigg|2\bigg({1+i\over 2}\bigg)^2
+2\bigg({1-i\over 2}\bigg)^2\bigg|^2 = 0.\label{GGG4}
\end{eqnarray}
The complete disappearance of $G^{(4)}(4\times 1)$ is a result of
orthogonality of $|2H, 2V\rangle$ with a NOON state. Thus the
projection measurement gives a null result.

Notice that even when $|\tau_1-\tau_2|>> T_c$, there is still
two-photon interference ($2\times 2$ case). So we expect that the
value in Eq.(\ref{GG4}) is smaller than the case when all four
photons are well separated in time ($1\times 4$ case), as in Fig.
3c. In this case, the state of the four photons is
\begin{eqnarray}
|\Phi^{\prime}\rangle = |H(\tau_1) H(\tau_2) V(\tau_3)
V(\tau_4)\rangle.\label{phi-tau1}
\end{eqnarray}
After expanding Eq.(\ref{G4hv}) by substituting Eq.(\ref{Ehv}), we
find that only terms of the form $\hat E_H\hat E_H\hat E_V\hat
E_V/4$ or its permutations have nonzero values operating on
$|\Phi^{\prime}\rangle$. There are 6 such terms. When each term
operates on $|\Phi^{\prime}\rangle$, say, $\hat E_H(t_1)\hat
E_H(t_2)\hat E_V(t_3)\hat E_V(t_4)|H(\tau_1) H(\tau_2) |V(\tau_3)
V(\tau_4)\rangle$, different permutations between $t_1=\tau_1,
t_2=\tau_2$ or between $t_3=\tau_3,t_4=\tau_4$ give $2\times2 =4$
contributions. Because of the distinguishability in time, all of
them are incoherent to each other and they add together by their
intensities (absolute value squares). So overall, we have
$G^{(4)}(1\times 4) = (1/4)^2\times 6\times 4 = 3/2$.

As can be seen, $G^{(4)}(1\times 4) > G^{(4)}(2\times 2)>
G^{(4)}(4\times 1)$. This means that when we adjust the path
difference between H and V, $G^{(4)}$ will drop from
$G^{(4)}(1\times 4)$ to $G^{(4)}(2\times 2)$ for the $2\times 2$
case and to $G^{(4)}(4\times 1)$ for the $4\times 1$ case. This is
somewhat similar to the Hong-Ou-Mandel effect but for four
photons, two from each side. Hence, the visibility of the
generalized Hong-Ou-Mandel dip is
\begin{eqnarray}
&&{\cal V} (2\times 2)= [G^{(4)}(1\times 4)-G^{(4)}(2\times 2)]/
G^{(4)}(1\times 4) \cr && \hskip .57in=(3/2-1)/(3/2) =
1/3\label{V}
\end{eqnarray}
for $2\times 2$ case. For the $4\times 1$ case, because
$G^{(4)}(4\times 1)=0$, we always have ${\cal V} (4\times 1)=1$.

Although the above pictures is straightforward and easy to
understand, it is not rigorous and it only applies to the two
extreme cases. In the following, we will use a multi-mode theory
of parametric down-conversion to accurately calculate the
four-fold coincidence rate and confirm the results above.

\subsection{Multi-mode treatment}

The multi-mode theory for a type-II parametric down-conversion
gives the quantum state in the form of \cite{rhee}
\begin{eqnarray}
|\Phi\rangle = |0\rangle + \eta |\Phi_{2}\rangle + {1\over
2}\eta^2 |\Phi_{4}\rangle ,\label{phi}
\end{eqnarray}
where
\begin{eqnarray}
|\Phi_{2}\rangle = \int
d\omega_1d\omega_2\Phi(\omega_1,\omega_2)\hat
a_H^{\dagger}(\omega_1) \hat
a_V^{\dagger}(\omega_2)|0\rangle,~~~~\label{phi2}
\end{eqnarray}
and
\begin{eqnarray}
&&|\Phi_{4}\rangle = \int
d\omega_1d\omega_2d\omega_1^{\prime}d\omega_2^{\prime}
\Phi(\omega_1,\omega_2)\Phi(\omega_1^{\prime},\omega_2^{\prime})\cr&&\hskip
0.8in \hat a_H^{\dagger}(\omega_1) \hat
a_V^{\dagger}(\omega_2)\hat a_H^{\dagger}(\omega_1^{\prime}) \hat
a_V^{\dagger}(\omega_2^{\prime})|0\rangle.~~~~~~~\label{phi4}
\end{eqnarray}
$|\Phi_{2}\rangle$ is a two-photon state and $|\Phi_{4}\rangle$ is
a four-photon state.

Now let us calculate the four-photon correlation function
$G^{(4)}$ in Eq.(\ref{G4hv}) with the state in Eq.(\ref{phi}).
After expanding the product in Eq.(\ref{G4hv}) with
Eq.(\ref{Ehv}), we find only the following combinations are
non-zero:
\begin{eqnarray}
&&\hat E_1(t_1)\hat E_2(t_2)\hat E_3(t_3)\hat E_4(t_4)|\Phi\rangle
\cr &&\hskip 0.3 in = [(HHVV-VVHH)+\cr &&\hskip 0.6 in +
i(VHVH+HVHV)- \cr &&\hskip 0.9 in -i(HVVH+VHHV)]|\Phi\rangle,
~~~~~~\label{EHV}
\end{eqnarray}
where $H=\hat E_H, V=\hat E_V$ and we keep the time ordering. For
a multi-mode state in Eq.(\ref{phi}), $\hat{E}_H, \hat{E}_V$ are
expressed in multi-mode as
\begin{eqnarray}
\hat E_{H,V}(t)= \int d\omega \hat{a}_{H,V}(\omega)e^{-i\omega
t},\label{EHV-m}
\end{eqnarray}
where $\hat{a}_{H,V}(\omega)$ is the annihilation operator
satisfying the commutation relation:
\begin{eqnarray}
[\hat{a}^{\dagger}(\omega), \hat{a}(\omega^{\prime})] =
\delta(\omega-\omega^{\prime}). \label{comm}
\end{eqnarray}

We can see that only the four-photon term in Eq.(\ref{phi}) will
contribute to Eq.(\ref{EHV}). The first term in Eq.(\ref{EHV}) can
be easily calculated as
\begin{eqnarray}
&& HHVV |\Phi\rangle  \cr &&\hskip 0.15 in = \eta^2\int d\omega_1
d\omega_2d\omega_1^{\prime}d\omega_2^{\prime}
\Phi(\omega_1,\omega_2) \Phi(\omega_1^{\prime},\omega_2^{\prime})
\times \cr &&\hskip 0.4 in \times
\Big(e^{-i\omega_1t_1-i\omega_1^{\prime}t_2}+
e^{-i\omega_1t_2-i\omega_1^{\prime}t_1}\Big)\times \cr &&\hskip .7
in \times \Big(e^{-i\omega_2t_3-i\omega_2^{\prime}t_4}+
e^{-i\omega_2t_4-i\omega_2^{\prime}t_3}\Big)|0\rangle \cr &&\hskip
.15 in = 2\eta^2 \Big [g(t_1,t_3)g(t_2,t_4) +
g(t_1,t_4)g(t_2,t_3)\Big]|0\rangle, ~~~~~~~\label{HHVV}
\end{eqnarray}
where
\begin{eqnarray}
g(t,t^{\prime}) = \int d\omega_1 d\omega_2 \Phi(\omega_1,\omega_2)
e^{-i\omega_1t-i\omega_2t^{\prime}}. ~~~~~~\label{g}
\end{eqnarray}
The rest of the terms in Eq.(\ref{EHV}) have the following forms
\begin{eqnarray}
&& VVHH |\Phi\rangle   \cr &&\hskip .15 in = 2\eta^2 \Big
[g(t_3,t_1)g(t_4,t_2) + g(t_3,t_2)g(t_4,t_1)\Big]|0\rangle,
~~~~~~~\label{VVHH}
\end{eqnarray}
\begin{eqnarray}
&& HVHV |\Phi\rangle   \cr &&\hskip .15 in = 2\eta^2 \Big
[g(t_1,t_2)g(t_3,t_4) + g(t_1,t_4)g(t_3,t_2)\Big]|0\rangle,
~~~~~~~\label{HVHV}
\end{eqnarray}
\begin{eqnarray}
&& VHVH |\Phi\rangle   \cr &&\hskip .15 in = 2\eta^2 \Big
[g(t_2,t_1)g(t_4,t_3) + g(t_2,t_3)g(t_4,t_1)\Big]|0\rangle,
~~~~~~~\label{VHVH}
\end{eqnarray}
\begin{eqnarray}
&& HVVH |\Phi\rangle   \cr &&\hskip .15 in = 2\eta^2 \Big
[g(t_1,t_2)g(t_4,t_3) + g(t_1,t_3)g(t_4,t_2)\Big]|0\rangle,
~~~~~~~\label{HVVH}
\end{eqnarray}
\begin{eqnarray}
&& VHHV |\Phi\rangle   \cr &&\hskip .15 in = 2\eta^2 \Big
[g(t_2,t_1)g(t_3,t_4) + g(t_2,t_4)g(t_3,t_1)\Big]|0\rangle,
~~~~~~~\label{VHHV}
\end{eqnarray}

Next, let us assume the symmetry of $\Phi(\omega_1,\omega_2)
=\Phi(\omega_2,\omega_1)$ when the delay between H and V is zero.
so that $g(t,t^{\prime})=g(t^{\prime},t)$. Then Eq.(\ref{EHV})
becomes
\begin{eqnarray}
&&\hat E_1(t_1)\hat E_2(t_2)\hat E_3(t_3)\hat E_4(t_4)|\Phi\rangle
\cr &&\hskip 0.2 in =4\eta^2 \Big [g(t_1,t_4)g(t_2,t_3) -
g(t_1,t_3)g(t_2,t_4)\Big]|0\rangle, ~~~~\label{EHV1}
\end{eqnarray}
and Eq.(\ref{G4hv}) becomes
\begin{eqnarray}
&&G^{(4)}(t_1,t_2,t_3,t_4) \cr &&\hskip 0.2 in = 16 |\eta|^4\Big
|g(t_1,t_4)g(t_2,t_3) - g(t_1,t_3)g(t_2,t_4)\Big|^2.
~~~~\label{G4f}
\end{eqnarray}
The four-photon coincidence probability is proportional to an
integral of $G^{(4)}$ with respect to all times:
\begin{eqnarray}
&&P_4 (0)\propto \int_{-\infty}^{+\infty} dt_1 dt_2 dt_3 dt_4
G^{(4)}(t_1,t_2,t_3,t_4) \cr &&\hskip 0.38 in = 32 |\eta|^4({\cal
A} -{\cal E}), ~~~~~~\label{P4}
\end{eqnarray}
where
\begin{eqnarray}
{\cal A} &=& \int_{-\infty}^{+\infty} dt_1 dt_2 dt_3 dt_4
|g(t_1,t_4)g(t_2,t_3)|^2 \cr& =& \int
d\omega_1d\omega_2d\omega_1^{\prime}d\omega_2^{\prime}
\Big|\Phi(\omega_1,\omega_2)
\Phi(\omega_1^{\prime},\omega_2^{\prime})\Big|^2,~~\label{cA}
\end{eqnarray}
\begin{eqnarray}
&&{\cal E} = \int
d\omega_1d\omega_2d\omega_1^{\prime}d\omega_2^{\prime}
\Phi(\omega_1,\omega_2)\times \cr &&\hskip 0.5 in \times
\Phi(\omega_1^{\prime},\omega_2^{\prime})\Phi^*(\omega_1,\omega_1^{\prime})
\Phi^*(\omega_2,\omega_2^{\prime}).~~~~~~~\label{cE}
\end{eqnarray}

When the delay between H and V is not zero, we may introduce a
delay factor of $e^{i\omega_2\Delta T}$ for the V-mode. Then
$g(t,t^{\prime})$ in Eq.(\ref{g}) becomes $\bar g(t,t^{\prime})
\equiv g(t,t^{\prime}-\Delta T)$. Notice that now $\bar
g(t,t^{\prime}) \ne \bar g(t^{\prime},t)$ for nonzero $\Delta T$.
Substituting $\bar g(t,t^{\prime})$ into
Eqs.(\ref{HHVV}--\ref{VHHV}) and carrying out the time integral in
Eq.(\ref{P4}), we obtain after some lengthy calculation
\begin{eqnarray}
&&P_4 (\Delta T)\propto  4 |\eta|^4\Big[12( {\cal A + E})+4{\cal
E}^{(2)}(\Delta T)+ \cr&&\hskip 0.7 in -8 {\cal E}_1^{(1)}(\Delta
T)-8{\cal E}_2^{(1)}(\Delta T)-8{\cal E}_3^{(1)}(\Delta
T)\cr&&\hskip 1.0 in +4{\cal A}^{(2)}(\Delta T)- 8{\cal
A}^{(1)}(\Delta T)\Big], ~~~~~~~~~~\label{P4DT}
\end{eqnarray}
where
\begin{eqnarray}
{\cal A}^{(1)}(\tau) =A(\tau)A(0) ~~{\rm and}~~ {\cal
A}^{(2)}(\tau) =A^2(\tau),\label{cAA}
\end{eqnarray}
with
\begin{eqnarray}
A(\tau) \equiv \int d\omega_1d\omega_2
\Big|\Phi(\omega_1,\omega_2) \Big|^2e^{i\omega_2\tau},\label{A}
\end{eqnarray}
and
\begin{eqnarray}
&&{\cal E}_1^{(1)}(\tau) = \int
d\omega_1d\omega_2d\omega_1^{\prime}d\omega_2^{\prime}
\Phi(\omega_1,\omega_2)\Phi(\omega_1^{\prime},\omega_2^{\prime})\times
\cr &&\hskip 0.8 in \times \Phi^*(\omega_1,\omega_2^{\prime})
\Phi^*(\omega_1^{\prime},\omega_1)e^{i(\omega_2-\omega_1)\tau},~~~~~~~\label{cEE1}
\end{eqnarray}
\begin{eqnarray}
&&{\cal E}_2^{(1)}(\tau) = \int
d\omega_1d\omega_2d\omega_1^{\prime}d\omega_2^{\prime}
\Phi^*(\omega_1,\omega_2)\Phi^*(\omega_1^{\prime},\omega_2^{\prime})\times
\cr &&\hskip 0.8 in \times \Phi(\omega_1,\omega_2^{\prime})
\Phi(\omega_1^{\prime},\omega_2)e^{i(\omega_2-\omega_1)\tau},~~~~~~~\label{cEE2}
\end{eqnarray}
\begin{eqnarray}
&&{\cal E}_3^{(1)}(\tau) = \int
d\omega_1d\omega_2d\omega_1^{\prime}d\omega_2^{\prime}
\Phi(\omega_1,\omega_2)\Phi(\omega_1^{\prime},\omega_2^{\prime})\times
\cr &&\hskip 0.8 in \times \Phi^*(\omega_1,\omega_1^{\prime})
\Phi^*(\omega_2,\omega_2^{\prime})e^{i(\omega_2-\omega_1)\tau},~~~~~~~\label{cEE3}
\end{eqnarray}
\begin{eqnarray}
&&{\cal E}_2(\tau) = \int
d\omega_1d\omega_2d\omega_1^{\prime}d\omega_2^{\prime}
\Phi(\omega_1,\omega_2)\Phi(\omega_1^{\prime},\omega_2^{\prime})
 \times \cr &&\hskip .9 in \times \Phi^*(\omega_1,\omega_2^{\prime})
\Phi(\omega_1^{\prime},\omega_2)\times \cr &&\hskip 1.3 in \times
e^{i(\omega_2-\omega_1)\tau} e^{i(\omega_2^{\prime}
-\omega_1^{\prime})\tau}. ~~~\label{cE2}
\end{eqnarray}
Notice that ${\cal A}^{(1)}(0) = {\cal A}^{(2)}(0)={\cal A}$ and
${\cal E}_1^{(1)}(0) = {\cal E}_2^{(1)}(0)= {\cal
E}_3^{(1)}(0)={\cal E}^{(2)}(0)={\cal E}$, and because of the
symmetry $\Phi(\omega_1,\omega_2) =\Phi(\omega_2,\omega_1)$, we
have ${\cal E}_1^{(1)*}(\tau) = {\cal E}_2^{(1)}(\tau)$ and ${\cal
E}_3^{(1)*}(\tau) = {\cal E}_3^{(1)}(\tau)$.

When the delay $\Delta T$ is much larger than the coherence time,
or the reciprocal of the bandwidth of $\Phi(\omega_1,\omega_2)$,
there is no overlap among all four photons. This corresponds to
the $1\times 4$ case and all the $\Delta T$-dependent terms in
Eq.(\ref{P4DT}) are zero. Hence we have $P_4$ at large delay as
\begin{eqnarray}
&&P_4 (\infty)\propto  48 |\eta|^4({\cal A + E}).
~~~~~~\label{P4inf}
\end{eqnarray}
The visibility of the generalized Hong-Ou-Mandel dip is then
\begin{eqnarray}
{\cal V} \equiv  {P_4 (\infty)-P_4 (0)\over P_4 (\infty)} = {{\cal
A}+5{\cal E}\over 3({\cal A+E})}. ~~~~~~\label{VV}
\end{eqnarray}
Note that ${\cal E}\le {\cal A}$ by Schwartz inequality. The
equality stands if and only if $\Phi(\omega_1,\omega_2)$ is
factorized as $\Phi(\omega_1,\omega_2) =
\phi(\omega_1)\phi(\omega_2)$. When ${\cal E}= 0$, we have
\begin{eqnarray}
{\cal V}=1/3, ~~~~~~\label{E=0}
\end{eqnarray}
which is exactly same as Eq.(\ref{V}) and corresponds to the
situation when the two pairs of down-converted photons are well
separated and independent of each other ($2\times 2$ case).  But
when ${\cal E}={\cal A}$, Eq.(\ref{VV}) becomes
\begin{eqnarray}
{\cal V}=1. ~~~~~~\label{E=A}
\end{eqnarray}
In this situation the two pairs of down-converted photons are
overlapped to form an indistinguishable four-photon entangled
state ($4\times 1$ case).

\section{NOON State Projection for demonstration of Four-Photon de Broglie Wavelength}

In this section, we will apply the NOON state projection
measurement to two pairs of photons in EPR polarization entangled
state of the form
\begin{eqnarray}
|\Phi^+\rangle = {1\over\sqrt{2}}\Big(|2H\rangle
+e^{2i\varphi}|2V\rangle\Big).\label{EPR}
\end{eqnarray}
When the pairs are overlapping and indistinguishable, it can be
shown that the four-photon state has the form same as in
Eq.(\ref{HOM2}) with the two modes denoting $\hat a_{H,V}$. If we
make the NOON state projection measurement, only the NOON state
part will contribute and the unwanted $|2,2\rangle$ state is
projected out because it is orthogonal to the NOON state in
Eq.(\ref{NOON}).

Similar to the situation in Section III, we need to discuss the
$2\times 2$ case and the $4\times 1$ case separately. We will
start with the simple pictures that have been proven to be
correct.

\subsection{Simple Pictures for Two Independent Pairs and Four
Entangled Photons}

We again label the times at which the two pairs are generated as
$\tau_1, \tau_2$, respectively. For the case in Fig.3a ($2\times
2$ case), we have $|\tau_1-\tau_2|>> T_c$ but for Fig.3b,
$|\tau_1-\tau_2|<< T_c$($4\times 1$). We can then write the
quantum state of the two pairs as
\begin{eqnarray}
|\Phi\rangle = |\Phi^+(\tau_1)\rangle \otimes
|\Phi^+(\tau_2)\rangle,\label{phi-tau2}
\end{eqnarray}
with $|\Phi^+\rangle$ given in Eq.(\ref{EPR}).

For the projection measurement in Fig.2 for $N=4$ case, the
four-photon coincidence is proportional to the four-photon
correlation function given in Eq.(\ref{G4hv})

Similar to the case in previous section, when applying the above
to the state in Eq.(\ref{phi-tau2}), we find six nonzero terms in
Eq.(\ref{G4hv}), that is, when (i) $t_1 =t_2=\tau_1,
t_3=t_4=\tau_2$ or $t_1 =t_2=\tau_2, t_3=t_4=\tau_1$; (ii) $t_1
=t_3=\tau_1, t_2=t_4=\tau_2$ or $t_1 =t_3=\tau_2, t_2=t_4=\tau_1$;
(iii) $t_1 =t_4=\tau_1, t_2=t_3=\tau_2$ or $t_1 =t_4=\tau_2,
t_2=t_3=\tau_1$. The case (i) can be calculated as
\begin{eqnarray}
&&\hat E_1(\tau_1)\hat E_2(\tau_1)\hat E_3(\tau_2)\hat
E_4(\tau_2)|\Phi\rangle \cr &&\hskip 0.2 in = \hat E_1(\tau_2)\hat
E_2(\tau_2)\hat E_3(\tau_1)\hat E_4(\tau_1)|\Phi\rangle \cr
&&\hskip 0.2 in = \hat E_1(\tau_1)\hat
E_2(\tau_1)|\Phi^+(\tau_1)\rangle\otimes \hat E_3(\tau_2)\hat
E_4(\tau_2)|\Phi^+(\tau_2)\rangle\cr &&\hskip 0.2 in \propto
(1-e^{i2\varphi})(1+e^{i2\varphi})|0\rangle;\label{EE1}
\end{eqnarray}
Similarly, the case (ii) gives
\begin{eqnarray}
&&\hat E_1(\tau_1)\hat E_2(\tau_2)\hat E_3(\tau_1)\hat
E_4(\tau_2)|\Phi\rangle \cr &&\hskip 0.3 in = \hat E_1(\tau_2)\hat
E_2(\tau_1)\hat E_3(\tau_2)\hat E_4(\tau_1)|\Phi\rangle\cr
&&\hskip 0.3 in \propto
(1+ie^{i2\varphi})(1+ie^{i2\varphi})|0\rangle;\label{EE2}
\end{eqnarray}
and the case (iii) gives
\begin{eqnarray}
&&\hat E_1(\tau_1)\hat E_2(\tau_2)\hat E_3(\tau_2)\hat
E_4(\tau_1)|\Phi\rangle \cr &&\hskip 0.3 in =\hat E_1(\tau_2)\hat
E_2(\tau_1)\hat E_3(\tau_1)\hat E_4(\tau_2)|\Phi\rangle \cr
&&\hskip 0.3 in \propto
(1-ie^{i2\varphi})(1-ie^{i2\varphi})|0\rangle.\label{EE3}
\end{eqnarray}

When the two pairs are separated, i.e, $|\tau_1-\tau_2|>> T_c$,
all three contributions are distinguishable and we add their
absolute values to give $G^{(4)}$:
\begin{eqnarray}
&&G^{(4)} \propto \big|(1-e^{i4\varphi})\big|^2
+\big|(1+ie^{i2\varphi})\big|^4 +\big|(1-ie^{i2\varphi})\big|^4
\cr &&\hskip 0.3in = 14 \bigg(1 - {3\over 7}\cos
4\varphi\bigg).\label{GG4hv}
\end{eqnarray}
On the other hand, when the two pairs overlap and become
indistinguishable, i.e., $|\tau_1-\tau_2|<< T_c$, we add the three
contributions in amplitudes:
\begin{eqnarray}
&&G^{(4)} \propto \Big|(1-e^{i4\varphi}) +(1+ie^{i2\varphi})^2
+(1-ie^{i2\varphi})^2\Big|^2 \cr &&\hskip 0.3in = 18 \big(1 - \cos
4\varphi\big).\label{GGG4hv}
\end{eqnarray}
In both cases, the four-photon coincidence measurement has
sinusoidal modulation with $4\varphi$ -- typical of 4-photon de
Broglie wave. But the first case only has 3/7= 42\% visibility but
the second case produces 100\% visibility, a result from a true
NOON state projection.

We next consider more rigorously the multi-mode theory.

\subsection{Multi-mode treatment}

There are many ways to produce two-photon entangled state in
Eq.(\ref{EPR}). The straightforward way is to use two collinear
degenerate type-I parametric down-conversion processes in series
but with their orientations orthogonal to each other. One of the
process gives the $|2H\rangle$ state while the other produces
$|2V\rangle$. If the two processes are pumped from a common
source, the final state will be in the form of Eq.(\ref{EPR}). In
the multi-mode theory, the quantum state for system up to the
four-photon has the form of
\begin{eqnarray}
&&|\Phi\rangle = |0\rangle + \eta_1 |\Phi_{H2}\rangle + \eta_2
|\Phi_{V2}\rangle + {1\over 2}\Big(\eta_1^2 |\Phi_{H4}\rangle +\cr
&&\hskip 0.6in +\eta_2^2 |\Phi_{V4}\rangle+2\eta_1\eta_2
|\Phi_{H2}\rangle |\Phi_{V2}\rangle\Big),\label{phi-I}
\end{eqnarray}
where
\begin{eqnarray}
|\Phi_{M2}\rangle = \int
d\omega_1d\omega_2\Phi(\omega_1,\omega_2)\hat
a_M^{\dagger}(\omega_1) \hat
a_M^{\dagger}(\omega_2)|0\rangle,~~~~\label{phi2-I}
\end{eqnarray}
and
\begin{eqnarray}
&&|\Phi_{M4}\rangle = \int
d\omega_1d\omega_2d\omega_1^{\prime}d\omega_2^{\prime}
\Phi(\omega_1,\omega_2)\Phi(\omega_1^{\prime},\omega_2^{\prime})\cr&&\hskip
0.8in \hat a_M^{\dagger}(\omega_1) \hat
a_M^{\dagger}(\omega_2)\hat a_M^{\dagger}(\omega_1^{\prime}) \hat
a_M^{\dagger}(\omega_2^{\prime})|0\rangle,~~~~~~~\label{phi4-I}
\end{eqnarray}
where $M=H,V$. $|\Phi_{M2}\rangle$ is a two-photon state and
$|\Phi_{M4}\rangle$ is a four-photon state.

Now let us calculate the four-photon correlation function
$G^{(4)}$ in Eq.(\ref{G4hv}) with the state in Eq.(\ref{phi-I}).
After expanding the product in Eq.(\ref{G4hv}) with
Eq.(\ref{Ehv}), we find only the following combinations are
non-zero:
\begin{eqnarray}
&&\hat E_1(t_1)\hat E_2(t_2)\hat E_3(t_3)\hat E_4(t_4)|\Phi\rangle
\cr &&\hskip 0.3 in = (HH-VV)(HH+VV) |\Phi\rangle+ \cr &&\hskip
0.6 in +i (VH-HV)(VH-HV)|\Phi\rangle, ~~~~~~\label{HV}
\end{eqnarray}
where $H=\hat E_H, V=\hat E_V$ and we keep the time ordering. The
first term in Eq.(\ref{HV}) can be easily calculated as
\begin{eqnarray}
&&(HH-VV)(HH+VV) |\Phi\rangle =  \cr &&\hskip 0.15 in =
\big(\eta_1^2HHHH|\Phi_{H4}\rangle -
\eta_2^2VVVV|\Phi_{V4}\rangle\big)/2 +\cr &&\hskip 0.5 in
+\eta_1\eta_2(HHVV-VVHH)|\Phi_{H2}\rangle |\Phi_{V2}\rangle\cr
&&\hskip .15 in = \big\{(\eta_1^2-\eta_2^2) F(t_1,t_2,t_3,t_4)/2 +
\cr &&\hskip 0.3 in +
2\eta_1\eta_2[g(t_1,t_2)g(t_3,t_4)-g(t_3,t_4)g(t_1,t_2)]\big\}|0\rangle\cr
&&\hskip .15 in = \big[(\eta_1^2-\eta_2^2)
F(t_1,t_2,t_3,t_4)/2\big]|0\rangle, ~~~~~~\label{HHVV0}
\end{eqnarray}
where $g(t,t^{\prime})$ is given in Eq.(\ref{g}) and
\begin{eqnarray}
&&F(t_1,t_2,t_3,t_4) \cr &&\hskip .15 in \equiv \langle 0| \hat
E_H(t_1)\hat E_H(t_2)\hat E_H(t_3)\hat E_H(t_4)|\Phi_{H4}\rangle
\cr &&\hskip .15 in = 8 \big[g(t_1,t_2)g(t_3,t_4)
+g(t_1,t_3)g(t_2,t_4)+\cr &&\hskip 1 in +
g(t_1,t_4)g(t_3,t_2)\big]. ~~~~~~\label{F}
\end{eqnarray}
Here we assumed the symmetry $\Phi(\omega_1,\omega_2)
=\Phi(\omega_2,\omega_1)$ for type-I PDC so that
$g(t,t^{\prime})=g(t^{\prime},t)$.

For the second term in Eq.(\ref{HV}), there are four contributions
after expanding the product and each can be calculated as in
Eq.(\ref{HHVV0}). We then have
\begin{eqnarray}
&&(VH-HV)(VH-HV) |\Phi\rangle \cr &&\hskip 0.15 in = \eta_1\eta_2
(VH-HV)(VH-HV) |\Phi_{H2}\rangle|\Phi_{V2}\rangle\cr &&\hskip 0.15
in = 8\eta_1\eta_2\big [g(t_1,t_3)g(t_2,t_4) -
g(t_1,t_4)g(t_2,t_3)\big]|0\rangle, ~~~~~~\label{VHHV0}
\end{eqnarray}
where we used $g(t,t^{\prime})=g(t^{\prime},t)$. Combining
Eqs.(\ref{HHVV0},\ref{VHHV0}), we have
\begin{eqnarray}
&&\hat E_1(t_1)\hat E_2(t_2)\hat E_3(t_3)\hat E_4(t_4)|\Phi\rangle
\cr &&\hskip 0.3 in = 4\big [b_1 (\eta_1^2-\eta_2^2) + 2 i b_2
\eta_1\eta_2\big]|0\rangle, ~~~~~~\label{EE}
\end{eqnarray}
with
\begin{eqnarray}
&&b_1  \equiv g(t_1,t_2)g(t_1,t_2) +g(t_1,t_3)g(t_2,t_4) +\cr
&&\hskip 1.4 in+ g(t_1,t_4)g(t_3,t_2) ~~~~~~\label{b1}
\end{eqnarray}
and
\begin{eqnarray}
&&b_2 \equiv g(t_1,t_3)g(t_2,t_4) - g(t_1,t_4)g(t_2,t_3).
~~~~~~\label{b2}
\end{eqnarray}
Next we assume $\eta_1=\eta$ and $\eta_2=\eta e^{2i\varphi}$,
where $\varphi$ is the phase difference between the vertical and
horizontal down-converted photons. Then Eq.(\ref{G4hv}) becomes
\begin{eqnarray}
&&G^{(4)}(t_1,t_2,t_3,t_4) \cr &&\hskip 0.3 in = 16 |\eta|^2\Big
|b_1 (1-e^{4i\varphi}) + 2 i b_2 e^{2i\varphi}\Big|^2  \cr
&&\hskip 0.3 in = 64 |\eta|^2\Big |b_2 - b_1 \sin 2\varphi
\Big|^2. ~~~~~~\label{G4f2}
\end{eqnarray}

After an integration of $G^{(4)}$ with respect to all time, we
obtain the four-photon coincidence probability as
\begin{eqnarray}
&&P_4\propto \int_{-\infty}^{+\infty} dt_1 dt_2 dt_3 dt_4
G^{(4)}(t_1,t_2,t_3,t_4) \cr &&\hskip 0.19 in = 64 |\eta|^2\Big
(B_2 + B_1 \sin^2 2\varphi - 2 B_{12} \sin 2\varphi \Big),
~~~~~~\label{P4n}
\end{eqnarray}
with
\begin{eqnarray}
&&B_1 = \int dt_1 dt_2 dt_3 dt_4 (|b_1|^2) = 3({\cal A} + 2{\cal
E}), ~~~~~~\label{B11}
\end{eqnarray}
\begin{eqnarray}
&&B_2 = \int dt_1 dt_2 dt_3 dt_4 (|b_2|^2) = 2({\cal A} - {\cal
E}), ~~~~~~\label{B22}
\end{eqnarray}
and
\begin{eqnarray}
&&B_{12} = \int dt_1 dt_2 dt_3 dt_4 (b_1b_2) = 0,
~~~~~~\label{B12}
\end{eqnarray}
where ${\cal A}, {\cal E}$ are given in Eqs.(\ref{cA},\ref{cE}),
respectively. Finally, we have
\begin{eqnarray}
&&P_4\propto 64|\eta|^2\Big ({\cal E} + 7{\cal A}/2 \Big)\Big(1-
{\cal V} \cos 4\varphi \Big), ~~~~~~\label{PP4n}
\end{eqnarray}
with
\begin{eqnarray}
{\cal V} \equiv  {3({\cal A}+2{\cal E})\over 7{\cal A} + 2{\cal
E}}. ~~~~~~\label{Vn}
\end{eqnarray}
When ${\cal E}= {\cal A}$, we have
\begin{eqnarray}
P_4\propto 18 \big(1- \cos 4\varphi \big), ~~~~~~\label{E=An}
\end{eqnarray}
which is exactly same as Eq.(\ref{GGG4hv}) and corresponds to the
situation when the two pairs overlap to form an indistinguishable
four-photon entangled state. However, when ${\cal E}=0$,
Eq.(\ref{PP4n}) becomes
\begin{eqnarray}
P_4\propto 14 \bigg(1- {3\over 7}\cos 4\varphi \bigg),
~~~~~~\label{E=0n}
\end{eqnarray}
which is same as Eq.(\ref{GG4hv}) and in this situation the two
pairs of down-converted photons are well separated and independent
of each other.

\section{Conclusion and Discussion}

From the above calculation, we find that the projection
measurement scheme that we designed in Section II has no
contribution from the non-NOON states due to orthogonal
projection. Thus it extracts out the contribution only from the
NOON state. The essence of the orthogonal projection is a
multi-photon interference where amplitudes instead of intensities
are added. The outcome of the measurement is highly dependent on
the multi-photon entanglement in the quantum state. From the
visibility, we can then quantitatively define the degree of
multi-photon entanglement. This is some issue we will discuss
elsewhere.

Although the scheme discussed in Section IV is somewhat similar to
the scheme by Walther et al \cite{wal}, there are some fundamental
differences in the two schemes. First, the starting quantum states
are different: we use two collinear type-I parametric
down-conversion processes to produce a state of the form in
Eq.(\ref{HOM2}) whereas the scheme by Walther et al starts with
two pairs of photons in EPR state. Secondly, the measurement
schemes are different: ours is a NOON state projection measurement
but the measurement by Walther et al is a special arrangement for
the cancellation of the unwanted terms. Of course, the end result
is the same: only the NOON state contributes to the measurement.
Thirdly, the phase variations in the scheme of Walther et al are
locally separated whereas in our scheme, the phases are all
together giving a true $4\varphi$ dependence.

From the construction process for the NOON state projection
measurement, we find that the method can be generalized to an
arbitrary projection measurement of N-photon superposition states
in Eq.(\ref{Ph}) since it relies on the factorization or finding
the roots of a polynomial equation:
\begin{eqnarray}
\sum_{n=0}^{N}c_n x^{N-n}y^n = 0 ,\label{Poly}
\end{eqnarray}
which always has solution.

In the discussion throughout the paper, we assumed that the
spatial modes of the two polarizations are perfectly matched. In a
real experiment, however, misalignment will result in imperfect
spatial mode match and lead to a reduced visibility in any
interference. The effect of misalignment can be incorporated in
our formulism by introducing a spatial factor of $e^{ik_Hx}$ or
$e^{ik_Vx}$ in Eq.(\ref{EHV-m})\cite{rhee}.  For simplicity, we
only consider a misalignment in propagation direction for the
spatial mismatch. More complicated cases of spatial mode match
will have similar result. Thus we can use one dimensional model
where $x$ is the coordinate of the detector along the direction of
${\bf k}_H-{\bf k}_V$. Four detectors will have four different
coordinates. The final result will be an integral over each
detector's size of $\Delta x$ (we assume all four detectors have
the same size) and this will lead to a non-unit visibility of $v =
{\rm sinc} (\pi \Delta x/L)$ for single photon interference at
each detector with $L \equiv \lambda/\Delta\theta$ as the single
photon interference fringe spacing ($\Delta\theta$ is the angle
between ${\bf k}_H$ and ${\bf k}_V$). Two-photon interference
visibility will be $v_2 = v^2$. With the spatial dependent factors
inserted in Eq.(\ref{EHV-m}), we may go through similar
calculation on the temporal integrals and carry out the spatial
integral. It is straightforward to show that the four-photon
interference visibility in Eq.(\ref{VV}) is changed to
\begin{eqnarray}
{\cal V} = {2v_2({\cal A}+3{\cal E})-v_2^2({\cal A+E})\over
3({\cal A+E})}, ~~~~~~\label{VVnn}
\end{eqnarray}
and Eq.(\ref{Vn}) to
\begin{eqnarray}
{\cal V} = {3({\cal A}+2{\cal E})v_2^2\over(6+v_2^2){\cal
A}+2{\cal E}(3-2v_2)}, ~~~~~~\label{Vnn}
\end{eqnarray}
where $v_2 = v^2 ={\rm sinc}^2 (\pi \Delta x/L)$ is the two-photon
interference visibility.

\begin{acknowledgments}
This work was funded by National Fundamental Research Program
(2001CB309300), the Innovation funds from Chinese Academy of
Sciences, and National Natural Science Foundation of China (Grant
No. 60121503). ZYO is also supported by the US National Science
Foundation under Grant No. 0245421.
\end{acknowledgments}

\end{document}